\documentstyle[aps, epsfig]{revtex}  % DON'T CHANGE
\newcommand{\lsim}{
\,\raisebox{0.35ex}{$<$}
\hspace{-1.7ex}\raisebox{-0.65ex}{$\sim$}\,
}
\begin{document}                % INITIALIZE - DONT CHANGE

\title{Effect of surface anisotropy on the hysteretic properties of a
magnetic particle}
\author{M. Dimian and H. Kachkachi\cite{CA}}
\address{Laboratoire de Magn\'{e}tisme et d'Optique, Universit\'e de Versailles St.
Quentin, \\
45 av. des Etats-Unis, 78035 Versailles, France}
\maketitle
\begin{abstract} 
We study the influence of surface anisotropy on the zero-temperature
hysteretic properties of a small single-domain 
magnetic particle, and give an estimation of the anisotropy   
constant for which deviations from the Stoner-Wohlfarth model are
observed. 
We consider a spherical particle with simple cubic
crystalline structure, a uniaxial anisotropy for core spins and
radial anisotropy on the surface, and compute the hysteresis loop by
solving the local Landau-Lifshitz equations for classical spin vectors.
We find that when the surface anisotropy constant is at least of the order of
the exchange coupling, large deviations are observed with
respect to the Stoner-Wohlfarth model in the hysteresis loop and thereby
the limit-of-metastability curve, due to the non-uniform cluster-wise
reversal of the magnetisation.
\end{abstract}
\section{Statement of the problem}
In this work, we deal with the effect of surface anisotropy on the
hysteretic properties (hysteresis loop and 
limit-of-metastability curve, the so-called SW astroid \cite{SW}) of a
single-domain spherical
particle of simple cubic crystalline
structure, uniaxial anisotropy in the core, and radial single-site
anisotropy for spins on the boundary.
We compute the hysteresis loop and thereby the critical field of this
particle by solving, at zero temperature, the   
local Landau-Lifshitz equation (LLE) derived from the classical anisotropic
Dirac-Heisenberg model in field (\ref{DH}).
Doing this for different values of the polar angle $\psi$ between the
applied field and the core easy axis, renders the critical field as a
function of $\psi$, that is the SW astroid.
Finally, our model Hamiltonian reads \cite{Kachkachi2} 
\begin{equation}\label{DH}
{\cal H}=-J\sum\limits_{\left\langle i,j\right\rangle }{\bf S}_{i}{\bf \cdot
S}_{j}-(g\mu _{B}){\bf H\cdot }\sum\limits_{i=1}^{{\cal N}}{\bf S}
_{i} 
-\sum\limits_{i}K_{i}({\bf S}_{i}{\bf \cdot e}_{i})^{2}, 
\end{equation}
where ${\bf S}_{i}$ is the unit spin vector on site $i,$ ${\bf H}$ is the
uniform magnetic field applied in a direction $\psi$ with
respect to the reference $z$ axis, ${\cal N}$ is the 
total number of spins; $J>0$ is the nearest-neighbour exchange
coupling; the last term in (\ref{DH}) is the uniaxial anisotropy
energy with easy axis ${\bf e}_{i}$ and constant $K_{i}>0$. 
Core spins have an easy axis along $z$ and constant $K_c$,
while surface spins have radial easy axes and constant $K_s$. 
In the sequel, we use the reduced parameters, $j\equiv J/K_c,\,
k_s\equiv K_s/K_c.$

The only free parameter of our model is the surface anisotropy
constant $k_s$. So upon varying it we estimate its value at 
which deviations from the (macroscopic) Stoner-Wohlfarth model are 
observed.
Details of the method used here and more results can be found in 
Ref.\cite{Kachkachi2}.

\section{Results and discussion}
\begin{figure}[h!]
\unitlength1cm 
\begin{picture}(9,8.5)(-1,-9.0)
\centerline{\epsfig{file=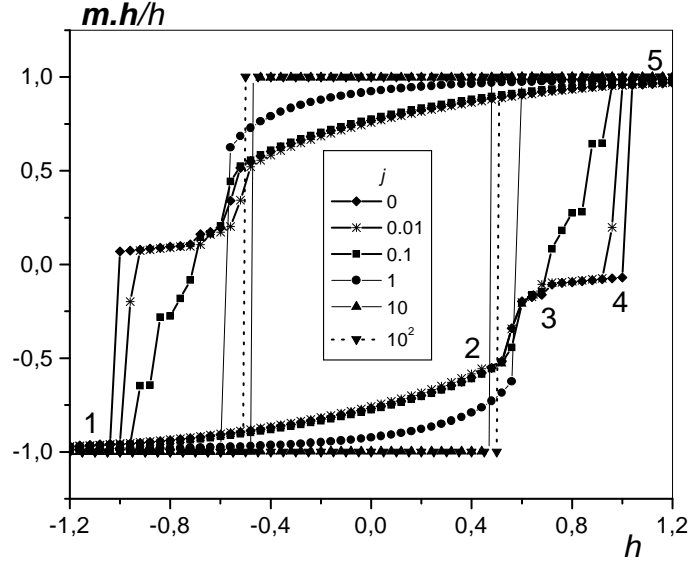,angle=-90,width=12.5cm}}
\end{picture}
\caption{\label{fig1_jap}
Hysteresis loop for $\psi =0$, $k_s=1$ and different values of
$j$. ${\cal N}=360$. 
}
\end{figure}

\begin{figure}[t]
\unitlength1cm 
\begin{picture}(9,8.5)(0,-9.0)
\centerline{\epsfig{file=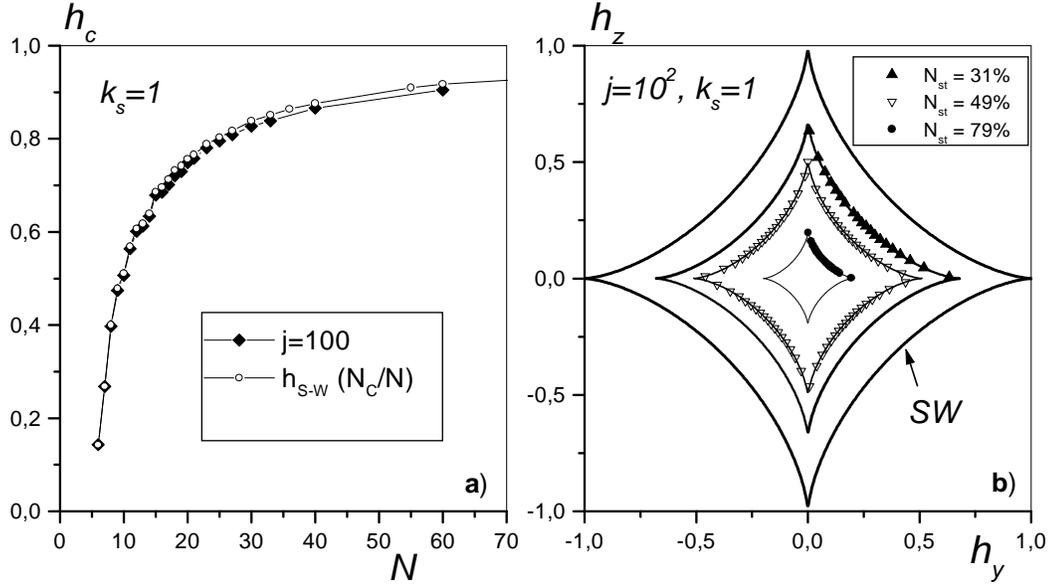,angle=-90,width=14.5cm}}
\end{picture}
\caption{\label{fig2_jap}
a) (in diamonds) Switching field for $k_s=1,\, j=10^2$ versus 
the particle's diameter $N$. (in circles) SW switching
field multiplied by $N_c/{\cal N}$.
b) Astroid for $k_s=1,\, j=10^2$ for different values of the
surface-to-volume ratio $N_{st}\equiv N_s/{\cal N}$.
}
\end{figure}

Fig.~\ref{fig1_jap} shows that when $k_s$ becomes comparable with $j$, the
competition between exchange coupling and surface anisotropy produces large
deviations from the SW model. 
Namely, the hysteresis loop exhibits 
multiple jumps, which can be attributed to the switching of different
spherical shells of spins starting from surface down to the centre. 
The hysteresis loop is then characterised by two field values: One
that marks the limit of metastability, called the {\it critical field} or the
saturation field, and the other that marks the magnetization
switching, and is called the {\it switching field} or the coercive field.
In Fig.~\ref{fig2_jap}a we present the variation with the particle's
diameter of the critical field \cite{footnote1} (in diamonds) obtained
from the numerical solution of the LLE for
$j=10^2$, and (in circles) the SW critical field multiplied by the
core-to-volume ratio, i.e. $N_c/{\cal N}$, analytically obtained upon
assuming infinite exchange coupling. This figure also shows that for
such a value of $k_s$, when the field is applied along the core easy
axis, the critical field increases with the particle's size, and that
for $k_s=1=10^{-2}j$ all hysteresis loops can be scaled with those
rendered by the SW model. 
%
%%%%%%%%%%%%%%%%%%%%%%%%%%%%%%%%%%%%%%%%%%%%%%%%%%%%%%%%%%%%%%%%%%%%%%%%%%%%%
%%%%%%%%%%%%%%%%%%%%%%%%%%%%%%%%%%%%%%%%%%%%%%%%%%%%%%%%%%%%%%%%%%%%%%%%%%%%%
%
%%%%%%%%%%%%%%%%%%%%%%%%%%%%%%%%%%%%%%%%%%%%%%%%%%%%%%%%%%%%%%%%%%%%%%%%%%%%%
%%%%%%%%%%%%%%%%%%%%%%%%%%%%%%%%%%%%%%%%%%%%%%%%%%%%%%%%%%%%%%%%%%%%%%%%%%%%%
%
On the other hand, these results clearly show that even in a large
particle there still
exist some magnetic inhomogeneities that lead to deviations from the SW
hypothesis of uniform rotation, as this is only rigorously realized in
an infinite system.
Moreover, Fig.~\ref{fig2_jap} shows that the critical field of a
spherical 
particle with $k_s=1$ can be obtained from the SW model through a
scaling with constant $N_c/{\cal N}$. One should also note that the
astroid for all particle sizes falls inside that of SW, see
Fig.~\ref{fig2_jap}b. Therefore, for 
$k_s/j\sim 0.01$ our model renders hysteresis loops and
limit-of-metastability curves that scale with the SW results for all
values of the angle $\psi$, the scaling constant being $N_c/{\cal N}$,
which is smaller than unity for a particle of any finite size. 
On the other hand, the critical field increases with the particle's
size and tends to the SW critical field in infinite systems.
%
%
%%%%%%%%%%%%%%%%%%%%%%%%%%%%%%%%%%%%%%%%%%%%%%%%%%%%%%%%%%%%%%%%%%%%%%%%%%%%%
\begin{figure}[t]
\unitlength1cm 
\begin{picture}(9,8.5)(0,-9.0)
\centerline{\epsfig{file=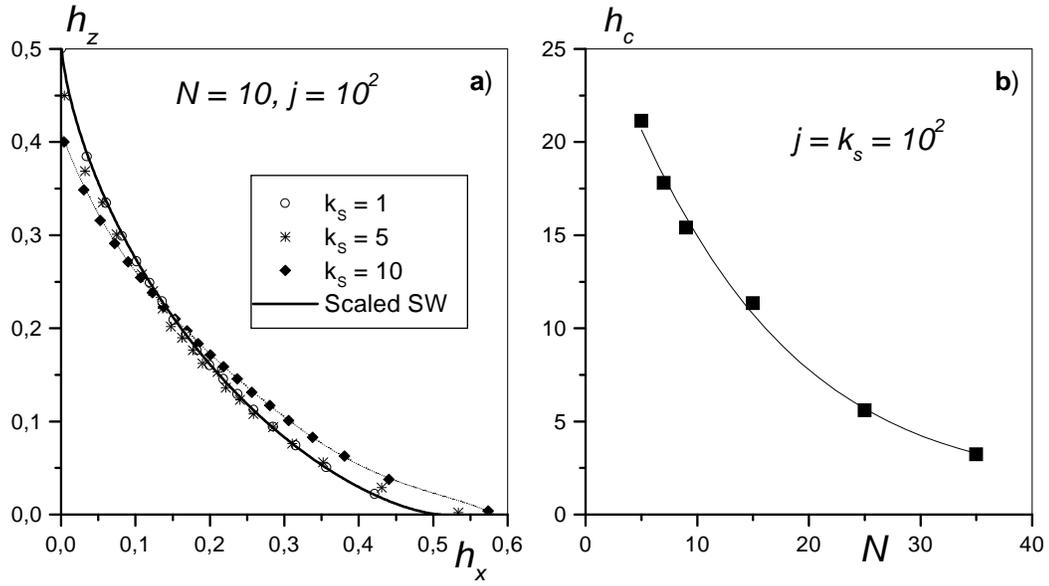,angle=-90,width=14.5cm}}
\end{picture}
\caption{\label{fig3_jap}
a) Astroid for $j=10^2$, ${\cal N}=360$ and different values of surface
anisotropy constant $k_{s}$. The full dark line is the SW
astroid scaled with $N_c/{\cal N}$, but the dotted line is only a
guide for the eye.
b) Switching field versus the particle's diameter $N$
for $\psi=0,\, j=k_s=10^2$.
}
\end{figure}
%%%%%%%%%%%%%%%%%%%%%%%%%%%%%%%%%%%%%%%%%%%%%%%%%%%%%%%%%%%%%%%%%%%%%%%%%%%%%
%
For larger values of $k_s/j$, but $k_s/j\lsim 0.2$, we still have some
kind of scaling but the corresponding constant now depends on
$\psi$. This is reflected by a deformation of the
limit-of-metastability curve. More precisely, as shown by
Fig.~\ref{fig3_jap}a, the latter is depressed in the core easy
direction and enhanced in the perpendicular direction. 
However, there is still only one jump in the hysteresis loop implying
that the magnetization reversal can be considered as uniform. On the
other hand, Fig.~\ref{fig3_jap}b shows that for $k_s/j\sim 1$ there
are two new features in comparison with the previous case (compare
Fig.~\ref{fig2_jap}a): the values of the switching field are much
higher, and more importantly, its behaviour as a function of the
particle's size is opposite to that of the previous case, as now the
field increases when the particle's size is decreased. For such high
values of $k_s$ ($K_s>>K_c$) surface spins are aligned along their
easy axes, and because of the strong exchange coupling they also drive
core spins in their switching process. Thus the smaller the particle
the larger the surface contribution, and the larger the field required
for a complete reversal of the particle's magnetization. 

%
%%%%%%%%%%%%%%%%%%%%%%%%%%%%%%%%%%%%%%%%%%%%%%%%%%%%%%%%%%%%%%%%%%%%%%%%%%%%%
\begin{figure}[h!]
\unitlength1cm 
\begin{picture}(9,8.5)(-1,-9.0)
\centerline{\epsfig{file=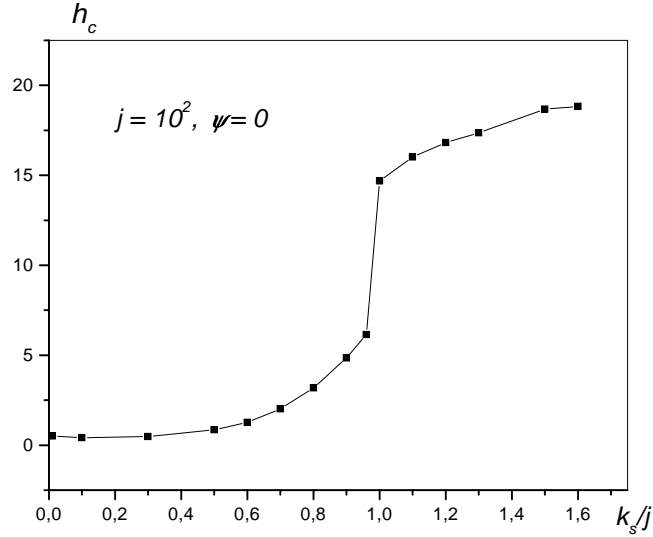,angle=-90,width=12cm}}
\end{picture}
\vspace{-1cm}
\caption{\label{fig4_jap} 
Switching field versus the surface anisotropy constant for $\psi =0$,
$j=10^2$, and $N=10$.
}
\end{figure}
%%%%%%%%%%%%%%%%%%%%%%%%%%%%%%%%%%%%%%%%%%%%%%%%%%%%%%%%%%%%%%%%%%%%%%%%%%%%%
%
In Fig.~\ref{fig4_jap} $h_c$ first slightly decreases for
$k_s/j\lsim 0.1$ and then increases, and when $k_s$ becomes of the
order of $j$ it jumps to large values. This plot illustrates the
different regimes discussed above according to the value of the ratio
$k_s/j$.
For an order of magnitude estimate of $K_s$ and the critical field
$H_c$ at the point $k_s/j=1$, consider a 4 nm spherical cobalt
particle for which $J\simeq$ 8 mev, $K_c\simeq 2.7\times 10^6$
erg/cm$^3$. Then, $K_s\sim 10$ erg/cm$^2$, and $H_c\sim 5$ T, when the
field is applied along the core easy axis.
\section{Conclusion}
Considering the fact that experiments on nanoparticles show that the
switching 
field does increase when the particle's size decreases (see e.g.
\cite{Chenetal} for cobalt particles), we may conclude
that the anisotropy constant $K_s$ is at least of the order of the
exchange coupling $J$, inasmuch as we can assume radial anisotropy on the
surface, as is usually done in the literature. Then, as discussed
above, for such values of $K_s$, large deviations are observed with 
respect to the SW model in the hysteresis loop and thereby
the limit-of-metastability curve, since in this case the magnetisation
reverses its direction in a non-uniform manner via a progressive switching of
spin clusters. So to deal with these features one has to resort to
microscopic approaches such as the one used in this work.
A work in progress applies the present technique to the
cubo-octahedral cobalt particles with a diameter of about 3nm recently
reported on in \cite{Jamet}.


\begin{references}
\bibitem{footnote1} For $k_s=1$, there is only one jump in
the hysteresis loop and thus the
critical field coincides with the switching field.
\end{references}
\end{document}